\documentclass[iop]{emulateapj}

\shorttitle{X-rays from BCDs}
\shortauthors{Kaaret, Schmitt, \& Gorski}

\begin{document}
\providecommand{\e}[1]{\ensuremath{\times 10^{#1}}}

\title{X-rays from Blue Compact Dwarf Galaxies}

\author{Philip Kaaret, Joseph Schmitt, and Mark Gorski}
\affil{Department of Physics and Astronomy, University of Iowa, Iowa City, IA 52242}

\begin{abstract}

We measured the X-ray fluxes from an optically-selected sample of blue compact dwarf galaxies (BCDs) with metallicities $<$0.07 and solar distances less than 15~Mpc.  Four X-ray point sources were observed in three galaxies, with five galaxies having no detectable X-ray emission.  Comparing X-ray luminosity and star formation rate, we find that the total X-ray luminosity of the sample is more than 10 times greater than expected if X-ray luminosity scales with star formation rate according to the relation found for normal-metallicity star-forming galaxies.  However, due to the low number of sources detected, one can exclude the hypothesis that the relation of the X-ray binaries to SFR in low-metalicity BCDs is identical to that in normal galaxies only at the 96.6\% confidence level.  It has recently been proposed that X-ray binaries were an important source of heating and reionization of the intergalactic medium at the epoch of reionization.  If BCDs are analogs to unevolved galaxies in the early universe, then enhanced X-ray binary production in BCDs would suggest an enhanced impact of X-ray binaries on the early thermal history of the universe.

\end{abstract}


\section{Introduction} 

Most bright X-ray sources within galaxies are compact objects. Massive, short-lived stars ($>8M_\sun$) leave behind compact objects that can form X-ray binaries, which are observed as point-like sources.  X-ray radiation can be used to study the recent past's star formation of a galaxy since a higher star-formation rate (SFR) will produce larger number of massive stars that then create more X-ray binaries \citep{Griffiths90,David92}.  Studies of nearby, near-solar-metallicity galaxies with active star formation show a well defined relation between X-ray luminosity and SFR \citep{Ranalli03,Grimm03,Kaaret08}.

There is significant uncertainty in using the locally measured correlation to model the early universe, e.g.\ \citep{Mirabel11}, because the properties of galaxies in the early universe were likely different from those of most nearby galaxies.  It has been suggested that blue compact dwarf (BCD) galaxies are local analogs of the small, metal-deficient galaxies found in the early universe \citep{Kunth00}.  Some BCDs are thought to be going through their first round of star formation, and most of their stellar population is due to a recent burst of star formation.  The X-ray emission of BCDs is dominated by point-like sources \citep{Thuan04}.

This paper reports on the relation between total X-ray luminosity and SFR for a sample of blue compact dwarf galaxies (BCDs) with metallicities $<$0.07 solar at distances less than 15~Mpc.  The X-ray observations and analysis of X-ray sources within each galaxy are described in section 2.  Section 3 describes estimation of the SFR for each galaxy based on infrared and H$\alpha$ luminosities reported in the literature.  We conclude, in section 4, with a discussion of the results and a comparison with the X-ray versus SFR correlation found in normal-metallicity galaxies.

\bigskip\bigskip

\begin{deluxetable*}{lccccccrr}
\tabletypesize{\scriptsize}
\tablecaption{Target Galaxies and X-ray Sources\label{xray}}
\tablewidth{0pt}
\tablehead{Galaxy  & Distance & d$_{25}$       & AGN Prob. & Exposure & Offset   & Counts    & Flux        & Luminosity  \\
                   & (Mpc)    & (arcsec)       &           & (ks)     & (arcsec) &           & (erg cm$^{-2}$ s$^{-1}$) 
                                                                                                           & (erg s$^{-1}$) }
\startdata
UGC 4483*          &  3.2     & 34 $\times$ 18 & 2.2\%     &  3.1     & ---      & ---       & $<$1.79E-14 & $<$2.2E+37  \\
VII Zw 403         &  4.3     & 24 $\times$ 13 & 0.1\%     & 10.2     & 14.1     & 289.9$^a$ & 1.57E-13    & 3.46E+38    \\
Mrk 209*           &  5.7     & 23 $\times$ 18 & 0.9\%     &  3.1     & ---      & ---       & $<$1.70E-14 & $<$6.6E+37  \\
DDO 68*            &  5.9     & 60 $\times$ 20 & 8.9\%     & 10.0     & 11.1     & ~15.8$^b$ & 8.38E-15    & 3.5E+37     \\
                   &          &                &           &          & 20.1     & ~14.9$^c$ & 7.90E-15    & 3.3E+37     \\
HS 1442+4250       &  10.0    & 26 $\times$ 6  & 0.8\%     &  5.2     & ---      & ---       & $<$1.02E-14 & $<$1.22E+38 \\
HS 0822+3542       &  10.9    & 9 $\times$ 4   & 0.3\%     &  5.1     & ---      & ---       & $<$1.11E-14 & $<$1.58E+38 \\
I Zw 18            &  12.6    & 11 $\times$ 9  & 0.1\%     & 40.1     & ~0.2     & 499.3$^d$ & 6.65E-14    & 1.26E+39    \\
UM 461*            &  13.4    & 12 $\times$ 9  & 2.1\%     & 34.9     & ---      & ---       & $<$1.54E-15 & $<$3.3E+37
\enddata
\tablecomments{The table includes: the galaxy name; distance; $d_{25}$ major and minor radii; the probability that a source at the measured flux level or upper limit is a background active galactic nucleus (AGN) taking into account the size of the $d_{25}$ ellipse of the galaxy; the observation exposure; the offset angular distance between the source and the center of the galaxy; the detected source counts; the flux in the 2--10~keV band of the detected source or the point source detection limit for a source with 10 counts; and the luminosity in the 2--10~keV band.  The names of detected sources are: $^{\rm a}$CXOU J112803.0+785953, $^{\rm b}$CXOU J095646.0+224929, $^{\rm c}$CXOU J095645.7+284920, $^{\rm d}$CXOU J093401.9+551428. Galaxies with an asterisk (*) were observed specifically for this study.} \bigskip
\end{deluxetable*}

\section{Observations and Analysis}

We searched the literature to construct a sample of BCDs at distances less than 15~Mpc and with metallicities $<$0.07 solar.  Specifically, our main source was the comprehensive review of BCDs of \citet{Kunth00}.  We also searched the Astrophysics Data System for papers on BCDs and examined the source lists in \citet{Hopkins02,Wu06}.  The resulting list, shown in Table~\ref{xray}, is not X-ray biased -- none of the galaxies was discovered via X-ray emission.  This sample is complete to the best of our knowledge, but was not constructed from an all sky survey.  However, the recent all sky infrared survey conducted by the {\it Wide-field Infrared Survey Explorer (WISE)} mission offers a means to construct a truly complete volume and luminosity limits sample of BCDs.  The fact that the first two new BCDs identified with {\it WISE} are at distances much larger than allowed in our sample supports the notion that our sample may be complete \citep{Griffith11}.

We acquired X-ray observations of all of the BCDs in the sample, see Table~\ref{xray}.  Four of galaxies were observed specifically for this program, UGC 4483, Mrk 209, DDO 68, and UM 461, and archival data were used for the rest.  All the observations were made with the Advanced CCD Imaging Spectrometer (ACIS) on the Chandra X-ray Observatory.  The Chandra Interactive Analysis of Observations (CIAO) package version 4.2 and the Chandra Calibration Database version 4.2.0 were used for analysis.

We searched for sources in the 0.3--8 keV band using the $wavdetect$ tool in CIAO with a significance threshold of $10^{-6}$.  Sources inside or with positional errors (typically $\sim 1 \arcsec$) overlapping the isophotal $d_{25}$ ellipse, calculated using parameters from the Hyperleda database, were accepted as being associated with the galaxy.  The detected sources are listed in Table \ref{xray}.  For galaxies with no detected sources, we estimated the flux that would have produced 10 counts in the image and use that to calculate a sensitivity limit.  To convert the source detections and sensitivity limits into fluxes, we assumed a power-law spectrum with a photon index of 1.7 and $N_H$ equal to Galactic absorption towards the host galaxy.  There was no evidence for diffuse emission in any of the galaxies.

We calculated the probability of finding an active galactic nucleus (AGN) at the flux level of the detected source or sensitivity limit within the $d_{25}$ ellipse of each galaxy using the flux distribution from \citet{Giacconi01}.  The probabilities are small except for DDO 68 with an 8.9\% probability of having a background AGN at a flux equal to or brighter than the dimmer source.  However, the sources in DDO 68 are a small fraction of the total luminosity and our results are not significantly affected by removal of one or both or these sources.  Indeed, in the discussion below we exclude these sources because they are below the luminosity threshold, $2\times10^{38} \rm \, erg s^{-1}$ \citep{} in the 2--10~keV band, used in the previous studies of the relation of X-ray binary population to star formation rate that we use for comparison \citep{Grimm03,Kaaret08}.

\begin{deluxetable*}{lccccc}
\tabletypesize{\scriptsize}
\tablecaption{Star Formation Rates\label{sfr}}
\tablewidth{0pt}
\tablehead{Galaxy          & SFR(24$\mu$m) & SFR(H$\alpha$, 24$\mu$m) & SFR(H$\alpha$) & SFR(FUV) & L$_{X}$(2-10~keV)  \\
                           & ($10^{-3}M_{\odot}$ yr$^{-1}$)  
                                           & ($10^{-3}M_{\odot}$ yr$^{-1}$)  
                                                                      & ($10^{-3}M_{\odot}$ yr$^{-1}$)  
                                                                                       & ($10^{-3}M_{\odot}$ yr$^{-1}$)  
                                                                                                  & ($10^{37}$ erg s$^{-1}$)  }
\startdata
UGC 4483$^{\rm b,l}$       & ~1.0          & ~~2.2                    & ~~2.7          &  ~~3.9   & ~$<$2.1  \\ 
VII Zw 403$^{\rm b,c,d,g}$ & ~7.5          & ~12.0                    & ~13.0          &  ~12.0   & ~34.6    \\ 
Mrk 209$^{\rm a,c,d,j}$    & 16.0          & ~36.0                    & ~41.0          &  ~20.0   & $<$10.0  \\ 
DDO 68$^{\rm a,c,d}$       & $<$2.4        & ~$<$1.7                  & ~~1.5          &  ~23.4   & ~~6.8    \\ 
HS 1442+4250$^{\rm e,h}$   & ~7.0          & ~11.0                    & ~10.0          &  ~21.4   & $<$13.7  \\
HS 0822+3542$^{\rm c,f,i}$ & ~3.7          & ~~6.4                    & ~~5.5          &  ~~2.9   & $<$11.0  \\
I Zw 18$^{\rm b,f}$        & ~9.4          & ~28.0                    & ~19.0          &  ~44.2   & 126.3    \\ 
UM 461$^{\rm k,l,m}$       & 38.0          & ~54.0                    & ~66.0          &  ~~8.3   & $<$6.0   \\  \hline
Total                      & 85.0          & 141.3                    & 158.7          &  136.1   & 167.7
\enddata
\tablecomments{References for H$alpha$ and 24$\mu$m data: $^{\rm a}$Dale et al. (2009), $^{\rm b}$Engelbracht et al. (2008), $^{\rm c}$Hunter \& Elmegreen (2004), $^{\rm d}$Hunter et al. (2010), $^{\rm e}$Kennicutt et al. (2008), $^{\rm f}$Kewley et al. (2007), $^{\rm g}$Lynds et al. (1998), $^{\rm h}$Popescu et al. (1999), $^{\rm i}$Rela\~no et al. (2007), $^{\rm j}$Schulte-Ladbeck et al. (2001), $^{\rm k}$Telles et al. (2001), $^{\rm l}$van Zee et al.\ (1998), $^{\rm m}$Wu et al. (2008).} \bigskip
\end{deluxetable*}

\section{Star Formation Rate}

SFR estimates are often uncertain, so we used four methods, see Table~\ref{sfr}.  The first was to calculate SFR using 24$\mu$m luminosities, $L(24\mu$m), found in the literature and applying equation 10 in Calzetti et al. (2010).  Of the many SFR relations based on $L(24\mu$m) considered by those authors, the one selected consistently gave the highest SFR.  Second, we found SFR estimates in the literature based on H$\alpha$ luminosity for each galaxy, except for HS 1442+4250 where the H$\beta$ luminosity was used.  Third, we used equation 17 in Calzetti et al. (2010) which combines 24$\mu$m and H$\alpha$ luminosity and corrects for the underestimation of SFR using $L(24\mu$m) alone.  Fourth, we used far ultraviolet (FUV) measurements obtained with the GALEX satellite.  The FUV SFRs were taken from \citet{Lee09}, except for I Zw 18 where we used the FUV luminosity from \citet{Gil07} and UM 461 where we used the results of \citet{Sargsyan09}.  When necessary, observational data in the relevant papers was converted into luminosity appropriate for the distances listed in Table \ref{xray}.  Where multiple references exist for the same parameter, the mean was taken.  


\begin{figure}
\centering
\includegraphics[width=0.5\textwidth]{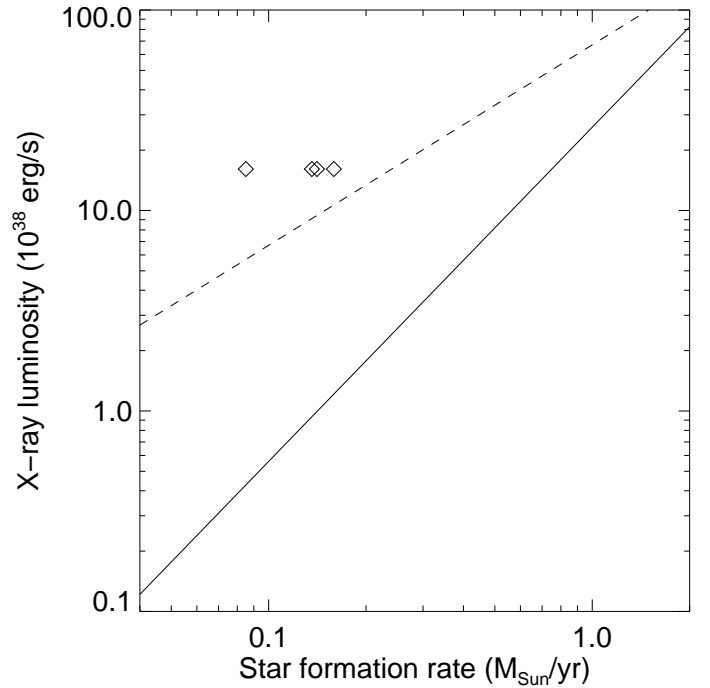} 
\caption{Sum of X-ray luminosities for source brighter than $2\e{38} \rm \, erg \, s^{-1}$ versus sum of SFR for the eight galaxies in our sample. From left to right, the diamonds represent SFR calculated from $L(24\mu$m), a combination of $L(24\mu$m) and $L(H\alpha$), $L(FUV)$, and $L(H\alpha$).  The solid line is the predicted relation between SFR and X-ray luminosity from studies of normal metallicity galaxies taking into account the statistics of the low number of sources found at low SFR  \citep{Gilfanov04}. The dashed line is a linear extrapolation of the relation measured at high SFR \citep{Grimm03}.\label{xvs}}
\end{figure}

\section{Results}

These galaxies are sufficiently small, and their star formation rates are sufficiently low, that fewer than one X-ray source is expected in any individual galaxy based on the measured SFRs for the galaxies and the relation between number of sources and SFR determined by \citet{Grimm03}.  Thus, we group all eight galaxies into one collection by summing their X-ray luminosities and SFRs.  Figure \ref{xvs} shows the relation between X-ray luminosity versus SFR.  Four points are plotted, one for each total SFR found in Table \ref{sfr}.  \citet{Grimm03} find that relation between X-ray luminosity and the SFR for normal-metallicity galaxies with SFR $\leq4.5 M_{\sun}$ yr$^{-1}$ is 
$$ L_{X} = (2.6\e{39} \rm \, erg \, s^{-1}) (SFR)^{5/3}. $$
\noindent The nonlinear form of the equation is not due to any change in the physical process involved in the interactions of SFR and X-ray luminosities.  Instead, it is due to the statistical properties of the small number of sources in galaxies with low SFRs \citep{Gilfanov04}.  This relation is superposed on Figure~\ref{xvs}.  The X-ray luminosity measurements of \citet{Grimm03} include point sources with luminosities down to $2\e{38} \rm \, erg \, s^{-1}$, thus, we have included only X-ray sources brighter than this in our total X-ray luminosity estimate of $1.6\e{39} \rm \, erg \, s^{-1}$.

The X-ray luminosity of the BCDs lies significantly above what would be predicted based on the known relation for normal-metallicity galaxies.  Taking the highest estimate of the SFR, that derived from $H\alpha$, we find that the measured X-ray luminosity is 13$\times$ that calculated from the previous equation.  We note that a linear extrapolation of the X-ray luminosity versus SFR measured at high SFR by \citet{Grimm03} is nearly consistent with our results, this is shown as a dashed line in Figure~\ref{xvs}.  However, the non-linear turnover at low SFR, and thus low numbers of X-ray sources, is a direct result of the application of Poisson statistics to the measured power-law distribution of source luminosities.  If the true relation is linear at low SFR, this would require that either the X-ray source populations do not obey Poisson statistics or the luminosity distribution function for low SFR galaxies differs significantly than that measured for high SFR galaxies.  \citet{Ranalli03} and \citet{Kaaret08} found a lower ratio of X-ray luminosity to SFR, at high SFR, than did \citet{Grimm03}.  This would further increase the discrepancy assuming that the reduction is applied to the non-linear curve of \citet{Grimm03} valid at low SFR.

However, due to the small number of sources, the result is not of high significance.  Using the relation between number of X-ray sources with luminosities above $2\e{38} \rm \, erg \, s^{-1}$ and SFR from \citet{Kaaret08}, our highest estimate of the SFR (inferred from H$\alpha$) leads to the prediction that 0.286 X-ray sources should be detected.  The probability distribution of the total number of sources follows a Poisson distribution.  Given a mean of 0.286, the probability to observe 2 or more sources is 3.4\%.  Thus, one can exclude that the hypothesis that the relation of the X-ray binaries to SFR in low-metallicity BCDs is identical to that in normal galaxies only at the 96.6\% confidence level.  Observations of a larger sample of low-metallicity BCDs is needed to further test this hypothesis.

Different SFR rate indicators are sensitive to star formation on different time scales.  H$\alpha$ is produced by the ionizing photons from hot, massive, and short-lived stars and traces the last $10^6$ years of star formation.  In contrast, infrared radiation is emission from dust heated by stars of a range of masses and averages roughly the past $10^8$ years of star formation.  The ``light-up'' times for high-mass X-ray binaries (HMXB) are of order $10^7$, corresponding to the lifetimes of the progenitor stars.  In a study H{\sc ii} galaxies that host young and intense star formation, \citet{Rosa09} found that the X-ray luminosity correlated with SFR according to the relation of \citet{Ranalli03} if the SFR was inferred from the IR luminosity.  In contrast, there was a deficit in the X-ray luminosity (in the 2--10~keV band) if H$\alpha$ was used as the SFR indicator.  In the H{\sc ii} galaxy sample, the H$\alpha$ SFR are much larger than the IR SFRs and signify a young and intense star formation event.  This lack of hard X-rays is expected since the star bursts are younger than the time needed for HMXB formation.  In our sample of BCDs, the H$\alpha$ SFRs are also larger than the IR SFRs, although by not as large a factor.  This suggests that the starburst is young, although perhaps not as young in the H{\sc ii} galaxies.  Using the IR SFR increases the discrepancy between the measured X-ray luminosity and that inferred based on the SFR.  The predicted number of X-ray sources is then 0.153 and the probability to observe two sources is 1.1\%.

\citet{Lee09} found that H$\alpha$ tends to underpredict the SFR when compared with FUV estimates.  The find that the ratio is a factor of 2 by SFR $\sim 0.003 M_{\odot} \rm \, yr^{-1}$, around the range of the lowest SFR galaxies considered here.  We examined the FUV SFR for our sample of BCDs and find that it is slightly lower than the H$\alpha$ SFR.  Thus, the effect found by \citet{Lee09} does not explain the overproduction of X-ray binaries in BCDs.

\citet{Mirabel11} have recently proposed that high-mass X-ray binaries in star-forming galaxies were an important source of heating and reionization of the intergalactic medium at the epoch of reionization.  If BCDs are analogs to the unevolved galaxies in the early universe, then enhanced X-ray binary production in BCDs would suggest an enhanced impact of X-ray binaries on the early thermal history of the universe over that predicted using X-ray versus SFR relations determined using near-solar metallicity galaxies.

\acknowledgments

We thank the referee for very useful comments.  We acknowledge the usage of the HyperLeda database (http://leda.univ-lyon1.fr) . This research has made use of data obtained from the Chandra Data Archive and software provided by the Chandra X-ray Center (CXC) in the application packages CIAO, ChIPS, and Sherpa.

\end{document}